\documentclass[final,journal]{IEEEtran}
\usepackage{amsmath}
\usepackage{mathtime}

\input{tcilatex}
\begin{document}

\title{Rashba Spin Interferometer}
\author{Ming-Hao Liu, Kuo-Wei Chen, Son-Hsien Chen, and Ching-Ray Chang \\
%EndAName
Department of Physics, National Taiwan University, Taipei 10617, Taiwan}
\maketitle

\begin{abstract}
A spin interferometer utilizing the Rashba effect is proposed. The novel
design is composed of a one-dimensional (1D) straight wire and a 1D
half-ring. By calculating the norm of the superposed wave function, we
derive analytical expressions to describe the spin interference spectrum as
a function of the Rashba coupling strength. Presented spin interference
results are identified to include (i) the quantum-mechanical $4\pi $
rotation effect, (ii) geometric effect, and (iii) Shubnikov-de Haas-like
beating effect.
\end{abstract}

\section{Introduction}

The Rashba spin-orbit coupling effect \cite{Rashba term}, originating from
the effective magnetic field generated by the electric field under the
relativistic transformation \cite{Winkler,Sakurai}, has stimulated plenty of
works in semiconductor spintronics \cite{SMspintronics}. For its
experimentally proved gate-voltage tunability \cite{VGtunibility InGaAs},
rotating the spins by Rashba effect (the Rashba spin precession \cite{MHL})
or even controlling the spin direction via the gate voltage \cite{Datta-Das}
are theoretically possible. It turns out that countless interesting spin
phenomena based on the Rashba effect have been proposed, including the spin
interference effect.

Nitta \textit{et. al.} first raised the issue of spin interferometer
considering a quantum ring with strong Rashba effect \cite{SI-Nitta}. Later
the analysis was extended from an ideal one-dimensional (1D) ring to a
two-dimensional ring under first quantization \cite{2D ring}, and second
quantization \cite{2D ring Nikolic}. Another design of the spin
interferometer using the square loop was also proposed \cite{SI-Nitta-SL},
and is even experimentally proved recently \cite{SI-Nitta-SLexp}.

In this paper, we propose another design for the Rashba spin interferometer,
which can be analytically solved. With obtained formulae, we identify the
spin interference into three categories: (i) quantum-mechanical $4\pi $
rotation effect, which states that a quantum state ket needs a rotation of $%
4\pi $ to bring it back, instead of $2\pi $ \cite{Sakurai,4pi rotation};
(ii) geometric effect, which arises from the geometry-dependent phase
difference; (iii) Shubnikov-de Haas (SdH)-like beating effect, which is the
most distinctive feature of our proposal.

\section{Theoretical Calculation}

Consider an interferometer device composed of a quantum wire attached by a
quantum half-ring, where the Rashba spin-orbit coupling is present. See Fig. %
\ref{fig1}(a). \FRAME{fbpFU}{2.7155in}{1.7391in}{0pt}{\Qcb{(a) Skematic
sketch of the spin interferometer device. Individual spin vectors calculated
on the half-ring and the wire are plotted in (b) and (c) with an injected $x$%
-polarized and $y$-polarized spins, respectively.}}{\Qlb{fig1}}{fig1.ps}{%
\special{language "Scientific Word";type "GRAPHIC";maintain-aspect-ratio
TRUE;display "USEDEF";valid_file "F";width 2.7155in;height 1.7391in;depth
0pt;original-width 5.4613in;original-height 3.5483in;cropleft "0";croptop
"1";cropright "1";cropbottom "0";filename '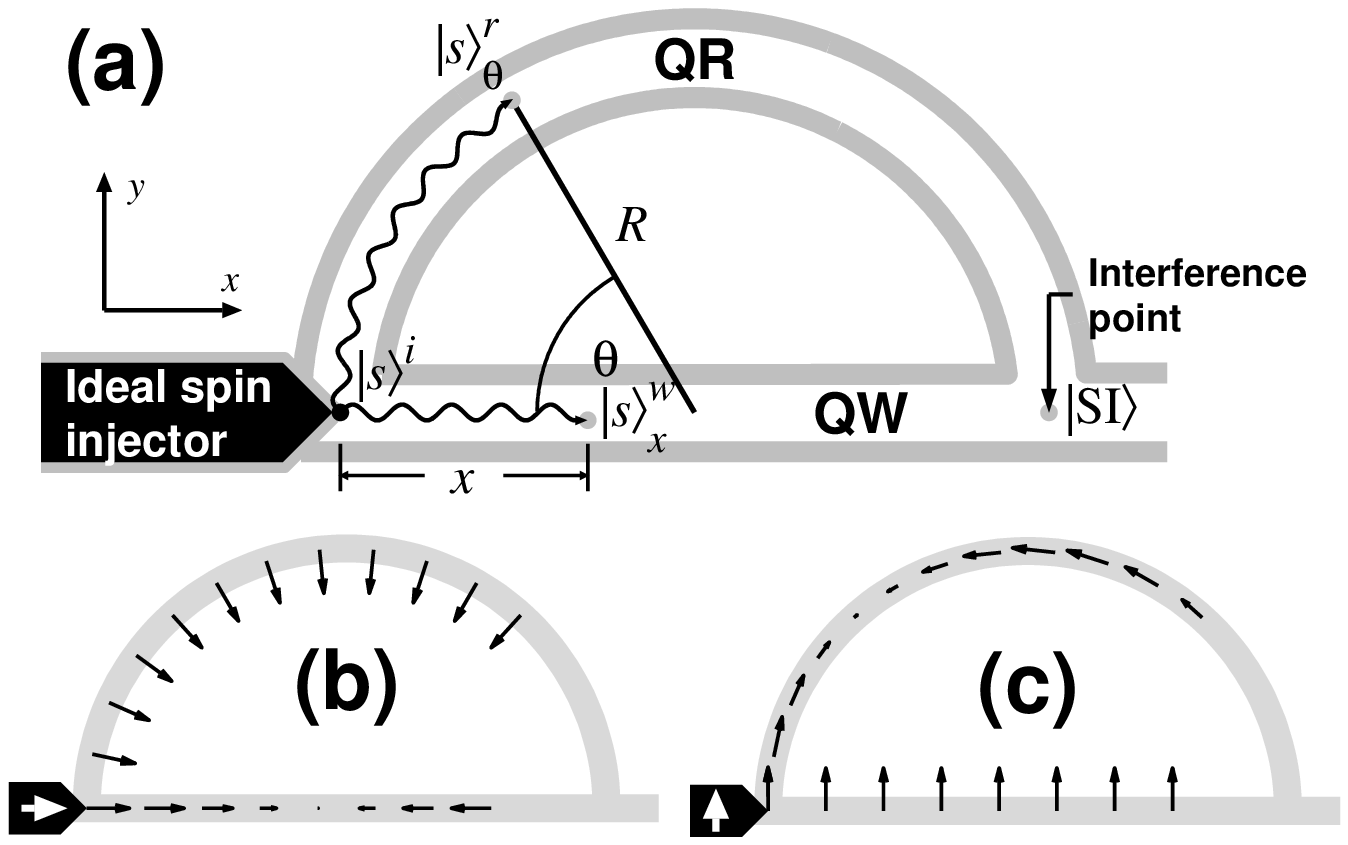';file-properties "XNPEU";}%
}An ideal spin injector is set up at one of the two connection points of the
ring-wire structure, while spin interference is expected to occur at the
other one. Assume that the injected spin, described by a state ket $%
\left\vert s\right\rangle ^{i}$, propagates through both the wire path,
yielding the state ket $\left\vert s\right\rangle _{x}^{w}$ at position $x$,
and the ring path, yielding $\left\vert s\right\rangle _{\theta }^{r}$ at
the angle $\theta $, with equal probability. Applying Ref. \cite{Nonuniform}%
, i.e., spatially translate the injected spin which is assumed to be
projected to the states at the Fermi level \cite{PSH}, we can write down the
individual state kets. For the straight wire, we have%
\begin{equation}
\left\vert s\right\rangle _{x}^{w}=e^{i\bar{k}_{w}x}\sum_{\sigma }e^{-i\frac{%
\sigma \Delta \theta _{w}}{2}}\left\langle \sigma ;\phi =0|s\right\rangle
^{i}\left\vert \sigma ;\phi =0\right\rangle   \label{ketw}
\end{equation}%
with $\bar{k}_{w}\equiv \left( k_{+}^{w}+k_{-}^{w}\right) /2$ and $\Delta
\theta _{w}\equiv 2m^{\star }\alpha _{w}x/\hbar ^{2}$, where $\alpha _{w}$
is the Rashba strength within the wire, and $k_{\pm }^{w}$ are the
spin-dependent wave vectors. In the present work, we will assume constant
Rashba strength in both the wire and the ring to facilitate the analytical
derivation. Spatial dependence of the Rashba coupling, if considered, can be
straightforwardly handled by the contour-integral method \cite{Nonuniform}.
For the half-ring, we similarly have%
\begin{equation}
\left\vert s\right\rangle _{\theta }^{r}=e^{i\bar{k}_{r}\pi \theta
}\sum_{\sigma }e^{-i\frac{\sigma \Delta \theta _{r}}{2}}\left\langle \psi
_{\sigma };\phi =\pi /2|s\right\rangle ^{i}\left\vert \psi _{\sigma };\phi
=\pi /2-\theta \right\rangle   \label{ketr}
\end{equation}%
with $\bar{k}_{r}\equiv \left( k_{+}^{r}+k_{-}^{r}\right) /2$ and $\Delta
\theta _{r}\equiv 2m^{\star }\alpha _{w}\pi \theta /\hbar ^{2}$, where we
denote the Rashba coupling strength within the ring as $\alpha _{r}$, which
in general may differ from $\alpha _{w}$. In both Eqs. (\ref{ketw}) and (\ref%
{ketr}), the well-known Rashba eigenspinor is given by $\left\vert \psi
_{\sigma };\phi \right\rangle \doteq \left( -ie^{i\phi },\sigma \right)
^{\dag }/\sqrt{2}$, where $\phi $ is the direction angle of the electron
wave vector. (Along $x$ axis we set $\phi =0$.)

Using Eqs. (\ref{ketw}) and (\ref{ketr}), one can depict the spin vectors
along individually the ring and the wire paths and clearly see the purpose
of our design of combining one straight wire and one half-ring. In Fig. \ref%
{fig1}(b), we inject an $x$-polarized spin, which propagates
precessionlessly along the half-ring and precessingly along the wire.
Conversely, if we inject a $y$-polarized spin, the precessing and
precessionless situations switch, as shown in Fig. \ref{fig1}(c). When
tuning the Rashba coupling strength (via the gate voltage \cite{VGtunibility
InGaAs}), which is equivalent to an effective magnetic field, the spin
coming out from the wire (ring) is rotated while that from the ring (wire)
is fixed in the case of $x$-polarized ($y$-polarized) injection of spin. We
therefore expect to see the effect of the quantum-mechanical rotation of the
state ket with $4\pi $ periods \cite{4pi rotation}, mapped from the external
magnetic field in vacuum to the effective magnetic field in solids.

We proceed by deriving the explicit form of $\left\vert SI\right\rangle $.
At the interference region, the spin-interfered state is superposed by $%
\left\vert SI\right\rangle =\left\vert s\right\rangle _{x=2R}^{w}+\left\vert
s\right\rangle _{\theta =\pi }^{w}$, regardless of normalization. Next we
define the dimensionless parameter%
\begin{equation}
\delta _{w\left( r\right) }\equiv 4m^{\star }\alpha _{w\left( r\right)
}R/\hbar ^{2}\text{,}  \label{delta}
\end{equation}%
which are responsible for the Rashba coupling and characterize the
precession angle of the injected spin $\Delta \theta _{w}=\delta _{w}$ and $%
\Delta \theta _{r}=\pi \delta _{r}/2$. Furthermore, under the assumption
that the spin is injected at $E_{F}$ so that we have $\bar{k}_{w\left(
r\right) }=\sqrt{2m^{\star }/\hbar ^{2}}\sqrt{E_{F}+\hbar ^{2}\delta
_{w\left( r\right) }^{2}/\left( 32m^{\star }R^{2}\right) }$, we can define%
\begin{equation}
\Delta _{w\left( r\right) }\equiv \bar{k}_{w\left( r\right) }R=\frac{\delta
_{w\left( r\right) }}{4}\sqrt{1+E/\delta _{w\left( r\right) }^{2}}\text{,}
\label{Delta}
\end{equation}%
where $E\equiv 32m^{\star }R^{2}E_{F}/\hbar ^{2}$ is dimensionless.
Therefore, we have the spin-interfered state%
\begin{eqnarray}
\left\vert SI\right\rangle &=&e^{i2\Delta _{w}}\sum_{\sigma }e^{-i\frac{%
\sigma \delta _{w}}{2}}\left\langle \psi _{\sigma };0|s\right\rangle
^{i}\left\vert \psi _{\sigma };0\right\rangle  \notag \\
&&+e^{i\pi \Delta _{r}}\sum_{\sigma }e^{-i\frac{\sigma \pi \delta _{r}}{4}%
}\left\langle \psi _{\sigma };\pi /2|s\right\rangle ^{i}\left\vert \psi
_{\sigma };-\pi /2\right\rangle \text{.}  \label{|SI>}
\end{eqnarray}

We now consider the specific cases of spin injection, namely, the $x$%
-polarized and the $y$-polarized injection of spin. Substituting $\left\vert
s\right\rangle ^{i}\doteq \left( _{1}^{1}\right) /\sqrt{2}$ into Eq. (\ref%
{|SI>}), one can, after some mathematical manipulation, obtain the
spin-interfered state ket $\left\vert SI;x\right\rangle $, yielding the spin
interference%
\begin{equation}
\left\langle SI;x|SI;x\right\rangle =2-2\sin \left( \delta _{w}/2\right)
\cos \left( \pi \Delta _{r}-2\Delta _{w}-\pi \delta _{r}/4\right) \text{.}
\label{SIx}
\end{equation}%
Similarly, with $\left\vert s\right\rangle ^{i}=\left( _{1}^{-i}\right) /%
\sqrt{2}$ we have%
\begin{equation}
\left\langle SI;y|SI;y\right\rangle =2-2\sin \left( \pi \delta _{r}/4\right)
\cos \left( \pi \Delta _{r}-2\Delta _{w}-\delta _{w}/2\right) \text{.}
\label{SIy}
\end{equation}

\section{Results and Discussion}

In the following we will use Eqs. (\ref{SIx}) and (\ref{SIy}), together with
Eqs. (\ref{delta}) and (\ref{Delta}), to demonstrate the $4\pi $ rotation
effect, the geometric effect, and the SdH-like beating effect in the
proposed interferometer device. Specifically, we consider InGaAs-based
materials with the electron effective mass $m^{\star }/m_{e}=0.03$ ($m_{e}$
is the electron rest mass) and the maximal Rashba coupling strength $0.03$ $%
\unit{eV}\unit{nm}$ \cite{Rashba in InGaAs}.

\subsection{$4\protect\pi $ Rotation Effect: Separate Control}

Consider the interferometer with the Rashba strength of the wire (or $\delta
_{w}$) tunable, while that of the ring (or $\delta _{r}$) fixed. The ring
radius is set $R=1$ $\unit{%
%TCIMACRO{\U{3bc}}%
%BeginExpansion
\mu%
%EndExpansion
m}$. We inject the spin at $E_{F}=5$ $\unit{eV}$, and tune $\alpha _{w}$
from $0$ to $0.03$ $\unit{eV}\unit{nm}$, corresponding to about $\delta
_{w}=0\rightarrow 15\pi $. Considering the $x$-polarized injection of spin,
the $4\pi $ rotation effect is clearly seen, as shown in Fig. \ref{fig_x}%
(a). \FRAME{ftbpFU}{2.7492in}{2.3583in}{0pt}{\Qcb{Spin interference spectrum
as a function of $\protect\delta _{w}$ by injecting an $x$-polarized spin.
In the separate control cases (a) and (c), $\protect\delta _{r}$ is fixed as
the maximum of $\protect\delta _{w}$ while in the on-board tuning cases (b)
and (d), $\protect\delta _{r}$ is set equal to $\protect\delta _{w}$. Labels
\textquotedblleft high energy\textquotedblright\ and \textquotedblleft low
energy\textquotedblright\ correspond to $E_{F}=5$ $\unit{eV}$ and $E_{F}=32$
m$\unit{eV}$, respectively.}}{\Qlb{fig_x}}{fig2.ps}{\special{language
"Scientific Word";type "GRAPHIC";maintain-aspect-ratio TRUE;display
"USEDEF";valid_file "F";width 2.7492in;height 2.3583in;depth
0pt;original-width 6.8044in;original-height 5.8271in;cropleft "0";croptop
"1";cropright "1";cropbottom "0";filename '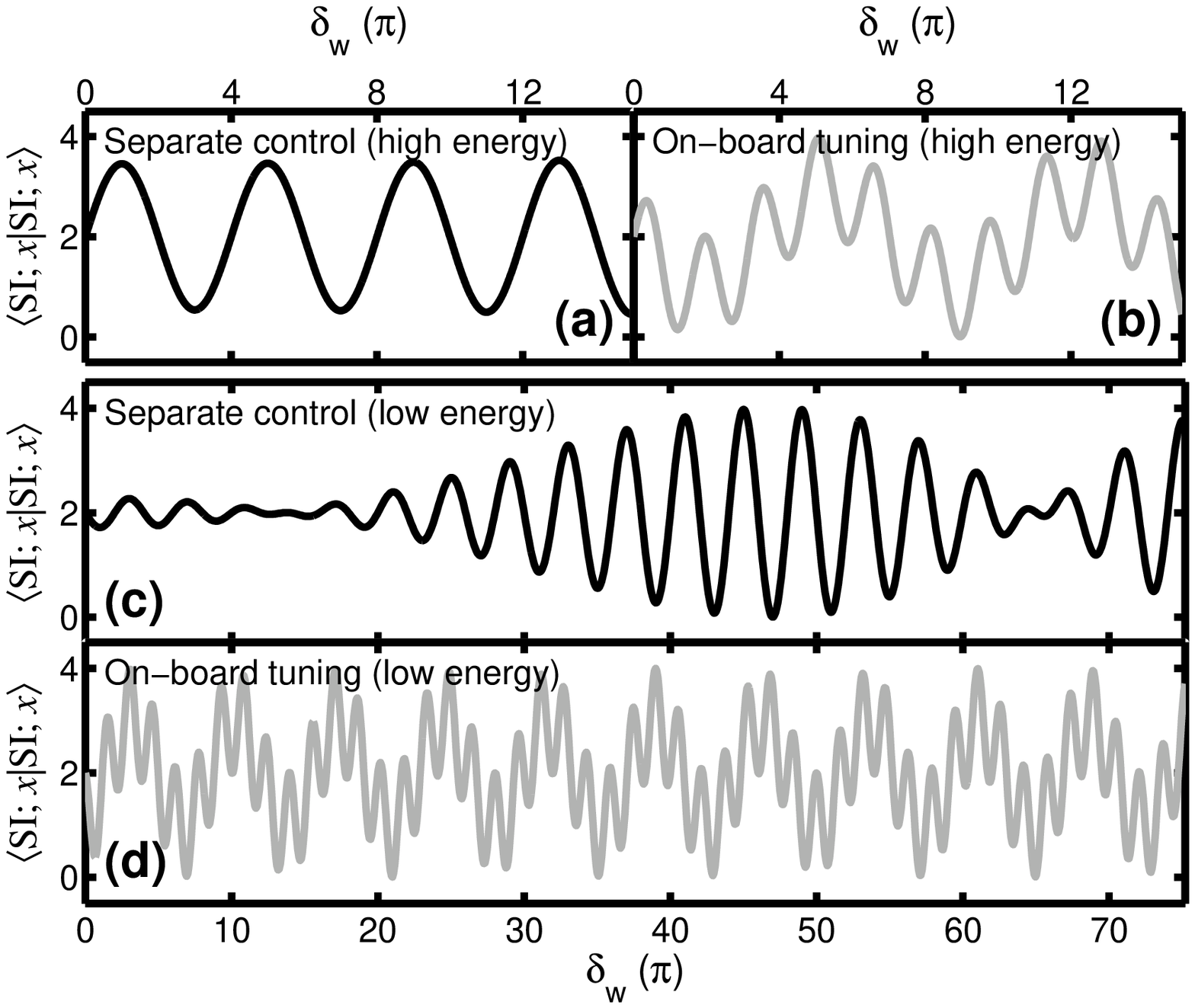';file-properties "XNPEU";}%
}Figure \ref{figy}(a) also shows the $4\pi $ rotation effect in the case of $%
y$-polarized injection of spin. \FRAME{fbpFU}{2.7492in}{2.3583in}{0pt}{\Qcb{%
Same as Fig. 2, except injecting a $y$-polarized spin.}}{\Qlb{figy}}{fig3.ps%
}{\special{language "Scientific Word";type "GRAPHIC";maintain-aspect-ratio
TRUE;display "USEDEF";valid_file "F";width 2.7492in;height 2.3583in;depth
0pt;original-width 6.8044in;original-height 5.8271in;cropleft "0";croptop
"1";cropright "1";cropbottom "0";filename '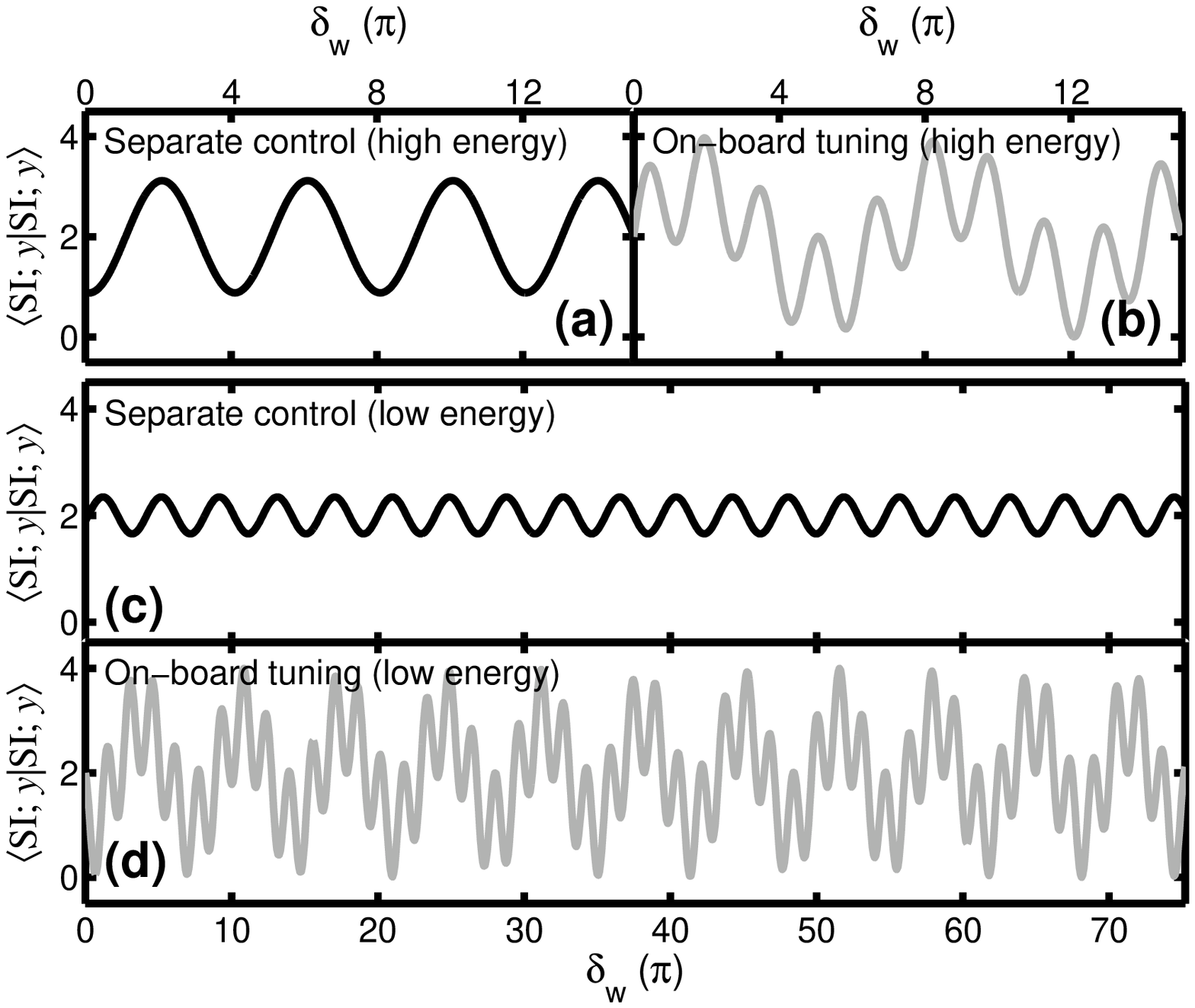';file-properties "XNPEU";}%
}In general, the period of $4\pi $ exists whenever we tune the Rashba
strength only on one side. Referring to Eq. (\ref{|SI>}) and noting that $%
\Delta _{w\left( r\right) }\approx $ $\sqrt{E}/4=$ constant for the given $%
E_{F}$ here [see Eq. (\ref{Delta})], one can see that the $4\pi $ period
stems from the partial wave $\left\vert s\right\rangle _{2R}^{w}$ passing
through the wire. When superposing $\left\vert s\right\rangle _{2R}^{w}$
with an arbitrary fixed state (not parallel to $\left\vert s\right\rangle
_{2R}^{w}$), the oscillation with the $4\pi $ period, which can never be
obtained by such a normalized state $\left\vert s\right\rangle _{2R}^{w}$
only, is revealed.

\subsection{Geometric Effect: On-Board Tuning}

We now tune $\delta _{r}$ and $\delta _{w}$ simultaneously ($\delta
_{r}=\delta _{w}$) via an on-board gate voltage, with other parameters
identical with the previous discussion. In this case the oscillation
behavior is completely changed, as can be seen in Figs. \ref{fig_x}(b) for
the $x$-polarized case and \ref{figy}(b) for the $y$-polarized case. Such an
interference effect originates from the difference in $\Delta \theta _{w}$
and $\Delta \theta _{r}$ shown in Eqs. (\ref{ketw}) and (\ref{ketr}), and is
therefore a geometric effect. Note that here since $\Delta \theta
_{r}-\Delta \theta _{w}=\left( \pi /4-1/2\right) \delta _{w}=\left( \pi
-2\right) \delta _{w}/2$, which is an irrational number of multiple of $%
\delta _{w}$, the state can no longer be returned to its original state, in
principle.

Note the distinct interference outcomes between Figs. \ref{fig_x}(a) and \ref%
{fig_x}(b), subject to exactly the same spin vectors along the individual
wire and ring paths [see Figs. \ref{fig1}(b)]! In this $x$-polarized case,
whether tuning $\delta _{r}$ or not, the spin vectors along the ring are
always fixed, since one of the spin channel is blocked. Specifically, the
expansion coefficients for the ring path are $\left\langle \psi _{\sigma
};\pi /2|s\right\rangle ^{i}=\delta _{1,\sigma }$ (the Kronecker delta
symbol defined by $\delta _{m,n}=0$ for $m\neq n$ and $\delta _{m,m}=1$) so
that $\left\vert s\right\rangle _{\pi }^{r}=e^{i\pi \left( \Delta
_{r}-\delta _{r}/4\right) }\left( _{1}^{-1}\right) /\sqrt{2}$ carries the
information of \emph{fixed} spin directions, no matter the phase is tuned or
not. However, superposing $\left\vert s\right\rangle _{2R}^{w}$ with $%
\left\vert s\right\rangle _{\pi }^{r}$ will eventually bring the
interference spectrum into $\left\langle SI|SI\right\rangle =2+2\func{Re}%
\left( \text{ }_{2R}^{w}\left\langle s|s\right\rangle _{\pi }^{r}\right) $,
which depends on whether $\left\vert s\right\rangle _{\pi }^{r}$ is varying
or not.

\subsection{Shubnikov-de Haas-Like Beating Effect}

Finally, we arrive at another feature of our spin interferometer, namely,
the beating effect. If we extend the range of $\delta _{w}$ by considering a
larger $R$, say $R=5$ $\unit{%
%TCIMACRO{\U{3bc}}%
%BeginExpansion
\mu%
%EndExpansion
m}$, and inject the spin at an lower energy, say $E_{F}=32$ m$\unit{eV}$,
one can clearly see the beating effect in the $x$-polarized case using the
separate control, as shown in Fig. \ref{fig_x}(c). On the contrary, such a
beating effect cannot be observed when injecting a $y$-polarized spin,
whether using the separate control or the on-board tuning.

Mathematically, the beating comes from the modulation of $\cos \left( \pi
\Delta _{r}-2\Delta _{w}-\pi \delta _{r}/4\right) $ on $\sin \left( \delta
_{w}/2\right) $ in Eq. (\ref{SIx}) for $\left\langle SI;x|SI;x\right\rangle $%
, while Eq. (\ref{SIy}) manifests that $\left\langle SI;y|SI;y\right\rangle $
does not exhibit any beating since $\cos \left( \pi \Delta _{r}-2\Delta
_{w}-\delta _{w}/2\right) $ is oscillating with $\delta _{w}$ but $\sin
\left( \pi \delta _{r}/4\right) $ is simply constant. Note also that such a
beating vanishes when injecting a high energy spin such that $\Delta
_{w}\approx $ constant as in the previous cases, since the cosine part in
Eq. (\ref{SIx}) becomes a constant. Physically, the beating effect is
similar to that in the SdH oscillation. In the ordinary SdH effect,
longitudinal resistance $\rho _{xx}$ oscillates with the increasing applied
perpendicular magnetic field $B_{\perp }$, which turns the
step-function-like density of states into (broadened) Landau levels. With
either spin-orbit coupling or an in-plane magnetic field $B_{\parallel }$,
each Landau level splits into two peaks due to spin, and the oscillation of $%
\rho _{xx}$ will compose of two close but different frequencies, leading to
the beating.

In the present Rashba spin interferometer, increasing $\delta _{w}$ by the
gate voltage is to increase the effective magnetic field, and we can thus
map $B_{\perp }$ and $\rho _{xx}$ in the ordinary SdH onto the Rashba field
strength (or $\delta _{w}$) and $\left\langle SI|SI\right\rangle $ in our
SdH-like beating effect, respectively. In the latter case, the phase factors
of the two spin components in the Rashba-tuned wave function (the state $%
\left\vert s\right\rangle _{2R}^{w}$ here) are of frequencies proportional
to their momentum $\hbar \left( \bar{k}_{w}\pm \Delta k\right) $ with $%
\Delta k$ being the wave vector difference proportional to $\alpha _{w}$.
When superposing $\left\vert s\right\rangle _{2R}^{w}$ with the fixed $%
\left\vert s\right\rangle _{\pi }^{r}$ state, the two closely spaced
frequencies lead to the beating, if the two spin components are equally
occupied. This is why we cannot observe the beating in the $y$-polarized
case. (One spin channel along the wire is blocked so that only one frequency
exists.)

\section{Conclusion}

In conclusion, we have shown the spin interference due to Rashba effect in
our proposed device. Obtained interference spectrum includes the 4$\pi $
rotation, the geometric, and the SdH-like beating effects. Presented results
suggest experimental measurement of simply the electric current through the
device versus the Rashba coupling strength, using either separate control or
on-board tuning. The former shows $4\pi $ periods while the latter does not,
when injecting higher energy spins with any polarization. SdH-like beating
is expected only when injecting $x$-polarized (actually, also valid for $z$%
-polarized) spins at low energy. The only nontrivial requirement is to tune $%
\delta $ by an enough wide range, which may require a larger ring, a
stronger spin-orbit coupling, or even a heavier effective mass. To maintain
the system as ballistic, meaning that the ring radius is of the order of
mircro-meter, estimation of $\delta $ from Eq. (\ref{delta}) shows that
InAs- or InGaAs-based materials seem to be the most feasible candidate.
GaAs-based material is less suggested, and SiGe/Si/SiGe symmetric quantum
wells are impossible in reality.

\section*{Acknowledgment}

We gratefully acknowledge financial support by the Republic of China
National Science Council Grant No. 95-2112-M-002-044-MY3.


\begin{thebibliography}{99}
\bibitem{Rashba term} E. I. Rashba, Sov. Phys. Solid State \textbf{2}, 1109
(1960); Yu. A. Bychkov and E. I. Rashba, JETP Lett. \textbf{39}, 78 (1984).

\bibitem{Winkler} R. Winkler, \textit{Spin-Orbit Coupling Effects in
Two-Dimensional Electron and Hole Systems} (Springer, Berlin, 2003).

\bibitem{Sakurai} J. J. Sakurai, \textit{Modern Quantum Mechanics}, revised
ed. (Addison-Wesley, New York, 1994).

\bibitem{SMspintronics} \textit{Semiconductor Spintronics and Quantum
Computation}, edited by D. D. Awschalom, D. Loss, and N. Samarth (Springer,
Berlin, 2002).

\bibitem{VGtunibility InGaAs} J. Nitta, T. Akazaki, H. Takayanagi, and T.
Enoki, Phys. Rev. Lett. \textbf{78}, 1335 (1997); T. Koga, J. Nitta, T.
Akazaki, and H. Takayanagi, Phys. Rev. Lett. \textbf{89}, 046801 (2002).

\bibitem{MHL} Ming-Hao Liu, Ching-Ray Chang, and Son-Hsien Chen, Phys. Rev.
B \textbf{71}, 153305 (2005); Ming-Hao Liu and Ching-Ray Chang, Phys. Rev. B 
\textbf{73}, 205301 (2006); Ming-Hao Liu and\ Ching-Ray Chang, J. Magn.
Magn. Mater. \textbf{304}, 97 (2006).

\bibitem{Datta-Das} S. Datta and B. Das, Appl. Phys. Lett. \textbf{56}, 665
(1990).

\bibitem{SI-Nitta} Junsaku Nitta, Frank E. Meijer, and Hideaki Takayanagi,
Appl. Phys. Lett. \textbf{75}, 695 (1999).

\bibitem{2D ring} Diego Frustaglia and Klaus Richter, Phys. Rev. B \textbf{69%
}, 235310 (2004).

\bibitem{2D ring Nikolic} Satofumi Souma and Branislav K. Nikoli\'{c}, Phys.
Rev. B \textbf{70}, 195346 (2004).

\bibitem{SI-Nitta-SL} Takaaki Koga, Junsaku Nitta, and Marc van Veenhuizen,
Phys. Rev. B \textbf{70}, 161302(R) (2004).

\bibitem{SI-Nitta-SLexp} Takaaki Koga, Yoshiaki Sekine, and Junsaku Nitta,
Phys. Rev. B \textbf{74}, 041302(R) (2006).

\bibitem{4pi rotation} S. A. Werner, R. Colella, A. W. Overhauser, and C. F.
Eagen, Phys. Rev. Lett. \textbf{35}, 1053 (1975).

\bibitem{Nonuniform} Ming-Hao Liu and Ching-Ray Chang, Phys. Rev. B \textbf{%
74}, 195314 (2006).

\bibitem{PSH} Ming-Hao Liu, Kuo-Wei Chen, Son-Hsien Chen, and Ching-Ray
Chang, Phys. Rev. B \textbf{74}, 235322 (2006).

\bibitem{Rashba in InGaAs} Y. Sato, T. Kita, S. Gozu, and S. Yamada, J.
Appl. Phys. \textbf{89}, 8017 (2001).
\end{thebibliography}
\end{document}